\documentclass[]{aa}
\usepackage{graphicx}
\usepackage{epsfig}
\begin{document}
\title{Approximate angular diameter distance in a locally
inhomogeneous universe with nonzero cosmological constant}
\author{M.Demianski \inst{1} \fnmsep \inst{2}, R. de Ritis \inst{3}
\fnmsep \inst{4}, A.A.Marino \inst{4}, E. Piedipalumbo\inst{3}
\fnmsep \inst{5}}
\offprints{E.Piedipalumbo, \\
\email{ester@na.infn.it}}
\institute{Institute for Theoretical Physics, University of Warsaw,
Warsaw, Poland
\and Department of Astronomy, Williams College, Williamstown, MA 01267
USA
\and Dipartimento di Scienze Fisiche, Universit\`{a} di Napoli
Federico II, Compl. Univ. Monte S. Angelo, 80126 Napoli, Italia
\and Istituto Nazionale di Fisica Nucleare, Sez. Napoli, Via
Cinthia, Compl. Univ. Monte S. Angelo, 80126 Napoli, Italia
\and Osservatorio Astronomico di Capodimonte, Via Moiariello,
16-80131 Napoli, Italia}

\date{Received / Accepted}

\titlerunning{Approximate angular diameter distance ..}
\authorrunning{M.Demianski et al.}

\abstract{We discuss the general and approximate angular diameter distance in the
Friedman-Robertson-Walker cosmological models with nonzero cosmological
constant. We modify the equation for the angular diameter distance by
taking into account the fact that locally the distribution of
matter is non homogeneous. We present exact solutions of this equation
in a few special cases. We propose an approximate analytic
solution of this equation which is simple enough and sufficiently
accurate to be useful in practical applications.
}
\maketitle

\section{Introduction}
Recent observations of the type Ia supernovae and CMB anisotropy
strongly indicate that the total matter-energy density of the universe
is now dominated by some kind of vacuum energy also called "dark
energy" or the cosmological constant $\Lambda$
(\cite{Perl97,perlal,rei&al98,Riess00}). The origin and nature of this
vacuum energy remains unknown. There are several review articles
providing thorough discussion of the history, interpretations and
problems connected with the vacuum energy and observational constrains
(\cite{zel67,weinberg2,car}).\\  The type Ia supernovae have been
already observed at redshifts $z>1$. It is   well known from galaxy
surveys that galaxies and clusters of galaxies up to a scale of
$\approx$ 1Gpc are distributed non homogeneously forming filaments,
walls and underdense voids. This indicates that on similar scales also
the dark matter is distributed non homogeneously. In this paper we
analyze the influence of local non homogeneities on the angular
diameter distance in a universe with non zero cosmological constant.
\\ The angular diameter distance in a locally non homogeneous universe was
discussed by Zeldovich ~(\cite{DZ65}) see also~(\cite{DS66}) and
(\cite{weinberg2,KHS97}).  Later Dyer\& Roeder (1972) used the so
called empty beam approximation to derive an equation for the angular
diameter distance; for more detailed references see also (\cite{SEF}),
(\cite{Kant98}), and (\cite{PRTP99}). We follow the Dyer
\& Roeder method to derive the equation for angular diameter distance in a
locally non homogeneous universe with a cosmological constant. In the
general case the equation for the angular diameter distance does not
have analytical solutions, it can be solved only numerically
(\cite{Kant20}). We have found an analytic approximate solution of
this equation, which is simple and accurate enough to be useful in
practical applications. This allowed us to find an approximate
dependence of the angular diameter distance on the basic cosmological
parameters.\\ The paper is organized as follows: we begin with the
general form of Sachs equations describing light propagation in an
arbitrary spacetime and using the empty beam approximation we derive
the equation for angular diameter distance. Then, we discuss the
angular diameter distance as observed from a location not at the
origin (z=0). Finally, after discussing some properties of the
analytical solutions of the equation for the angular diameter distance
in a locally  non homogeneous universe, we propose an analytic
approximation solution of this equation valid in a wide redshift
interval (0, 10). We have been motivated by the great advances in
observing further and further objects and the construction of very
powerful telescopes that provide the possibility to observe
gravitational lensing by clusters of galaxies, and supernovae at large
distances. It is therefore necessary to develop a more accurate
formalism to describe the distance redshift relation and gravitational
lensing by high--redshift objects. To achieve this goal we use
cosmological models with realistic values of the basic cosmological
parameters such as the Hubble constant and the average matter-energy
density. In concluding remarks we summarize our results  and discuss
some perspectives .
\section{Propagation of light and local non
homogeneities}
\subsection{General considerations} Let us consider
a beam of light emanating from a source S in an arbitrary
spacetime described by the metric tensor $g_{\alpha \beta}$. The
light rays propagate along a null surface $\Sigma$, which is
determined by  the eikonal equation \begin{equation}
 g^{\alpha\beta}\Sigma,_{\alpha}\Sigma,_{\beta}=0.
\label{eq:eiko}
\end{equation}
A light ray is identified with a null geodesic on $\Sigma$ with the tangent
vector $k_{\alpha}=-\Sigma,_{\alpha}$. The light rays in the beam can be
described by $x^{\alpha}=x^{\alpha}(v, y^{a})$, where $v$ is an affine
parameter, and $y^{a}$ (a=1, 2, 3) are three  parameters specifying
different rays. The vector field tangent to the light ray congruence,
$k^{\alpha}=\displaystyle{{dx^{\alpha}\over dv}}=-\Sigma,_{\alpha}$,
determines two optical scalars, the expansion $\theta$ and the shear
$\sigma$, which are defined by \begin{equation}
\theta\equiv{1\over 2}{k^{\alpha}}_{;\alpha},\;
\ \ \ \   \sigma\equiv k_{\alpha;\beta}{\bar m}^{\alpha}{\bar
m}^{\beta},
\label{eq:opscal1}
\end{equation}
where ${\bar
m}^{\alpha}={1\over\sqrt{2}}(\xi^{\alpha}-i\eta^{\alpha})$ is   a
complex vector spanning the spacelike 2-space (the screen space)
orthogonal to $k^{\alpha}$ $(k^{\alpha}{\bar m}_{\alpha}=0)$. Since
$k_{\alpha}=-\Sigma,_{\alpha}$ the vorticity connected with the light
beam is zero    in all our considerations; therefore in this case
$\sigma$ and $\theta$ fully    characterize the congruence. These two
optical scalars satisfy the Sachs (\cite{Sachs})   propagation
equations
\begin{equation}
\dot{\theta} +\theta^2 + |\sigma |^2 = -
\frac{1}{2} R_{\alpha \beta} k^{\alpha} k^{\beta},
\label{eq:raysachs1}
\end{equation}
\begin{equation}
\dot{\sigma} + 2\theta \sigma = -
\frac{1}{2} C_{\alpha \beta \gamma \delta} {\bar m}^{\alpha}
k^{\beta}{\bar m}^{\gamma} k^{\delta},
\label{eq:raysachs2}
\end{equation}
where  dot denotes the derivative with respect to $v$,
$R_{\alpha\beta}$ is the Ricci tensor, and $C_{\alpha \beta
\gamma\delta}$ is the Weyl tensor. Eqs. (\ref{eq:raysachs1}) and
(\ref{eq:raysachs2}) follow from   the Ricci identity. The optical scalars
$\theta$ and $\sigma$ describe the relative rate of change of an
infinitesimal area A of the cross section of the beam of light rays and
its distortion. The expansion $\theta$ is related to the relative   change
of an infinitesimal surface area $A$ of the beam's cross section by
\begin{equation}
\theta={1\over 2} {d\ln{A}\over dv}.
\label{eq:opscal2}
\end{equation}
Let us use these equations to study the propagation of   light in the
Friedman-Robertson-Walker (FRW) spacetime. The FRW spacetime   is
conformally flat, so that $C_{\alpha\beta\gamma\delta}=0$. From Eq.
(\ref{eq:raysachs2}) it follows that, in the FRW spacetime if the
shear of the null  ray congruence is initially zero, then it always
vanishes. Therefore, assuming that the light beam emanating from the
source S has vanishing shear, we can disregard the shear parameter
altogether (empty beam approximation). Using Eq. (\ref{eq:opscal2}) we
can rewrite equation (\ref{eq:raysachs1}) in the form
\begin{equation}
\ddot{\sqrt{A}}+{1\over 2}R_{\alpha\beta}k^{\alpha}k^{\beta}\sqrt{A}=0.
\label{eq:raysachs3}
\end{equation}
An observer moving with the 4-velocity vector $u^{\alpha}$ will associate
with  the light-ray  a circular frequency $\omega=cu^{\alpha}k_{\alpha}$.
Different observers assign different frequencies to the same light ray.
The shift of frequency, as measured by an observer   comoving with the
source and an arbitrary observer, is   related to their relative velocity
or the redshift $z$ by
\begin{equation}
1+z={\omega \over\omega_{o}}={c\over\omega_{o}}u_{\alpha}k^{\alpha},
\label{eq:redshift1}
\end{equation}
where $\omega$ is the frequency measured by a comoving observer and
$\omega_0$   by an arbitrary observer. Differentiating this equation
with respect to the affine parameter $\it{v}$, we obtain
\begin{equation}
{dz\over dv}={c\over \omega_{o}}k^{\alpha}k^{\beta}u_{\alpha;\beta}.
\label{eq:dzdv}
\end{equation}
Since the angular diameter distance D is proportional to $\sqrt{A}$ we
can rewrite Eq.(\ref{eq:raysachs3}) using D instead of $\sqrt{A}$ and,
at the same time,  replace  the affine parameter $\it{v}$ by the
redshift $z$; we obtain   \begin{equation}   \left ( \frac{dz}{dv}
\right )^2 \frac{d^2 D}{dz^2} + \left (   \frac{d^2z}{dv^2} \right )
\frac{dD}{dz} +\frac{4\pi G}{c^{4}} T_{\alpha   \beta} k^{\alpha} k^{\beta} D = 0,
\label{eq:angdiam2}
\end{equation}
where we have used the Einstein equations to replace the Ricci
tensor by the energy-momentum tensor. A solution
of~Eq.~(\ref{eq:angdiam2}) is related to the angular diameter
distance if it satisfies the following initial conditions:
\begin{eqnarray}\label{eq:cauchy}   &&D(z)\arrowvert_{z = 0} = 0,
\nonumber\\ && \\ &&\frac{dD(z)}{dz}   \arrowvert_{z = 0} =
\frac{c}{H_0}.\nonumber
\end{eqnarray}
\subsection{The angular diameter distance between two objects at
different redshifts}

In order to use solutions of the Eq. (\ref{eq:angdiam2}) to describe
gravitational lenses we have to introduce the angular diameter
distance between the source and the lens $D(z_{l}, z_{s})$, where
$z_{l}$ and $z_{s}$   denote correspondingly the redshifts of the lens
and the source.  Let $D(z_{1},z_{2})$ denote the angular diameter
distance between a fictitious observer at $z_{1}$ and a source at
$z_{2}$; of course, $D(0,z)=D(z)$. Suppose that we  know the general
solution  of Eq.~(\ref{eq:angdiam2}) for $D(z)$, which satisfies the
initial conditions (\ref{eq:cauchy});    let us define $D(z_{1},z)$ by
\begin{equation}
D(z_1, z) = \frac{c}{H_0}(1+z_1) D(z_1) D(z)   \left|
\int_{z_1}^{z}{\frac{dz'}{D^2(z') g(z')}}\right| \ ,   \label{eq:dzunoz}
\end{equation}
subject to the conditions
\begin{eqnarray}   &&D(z_1,
z) \arrowvert_{z= z_1} = 0, \nonumber \\ &&\\ &&\frac{d}{dz}   D(z_1, z)
\arrowvert_{z = z_1} = {\rm sign}(z - z_1)
\frac{c}{H_0}\frac{1+z_1}{g(z_1)}. \nonumber
\label{eq:rz1z2}
\end{eqnarray}
We see that $D(z_{1},z)$ satisfies
Eq.~(\ref{eq:angdiam2}), if the   function $g(z)$ is a solution of the
equation
\begin{eqnarray}
\frac{d}{dz} \ln{g(z)} =   {\frac{{\large
d^2z}}{{\large dv^2}}\over \left ( \frac{{\large   dz}}{{\large dv}}
\right )^2},   \label{eq:gzeq}
\end{eqnarray}
so that
\begin{equation} g(z) = g_0   \exp{\int{\frac{\frac{d^2z'}{dv^2}}{\left (
\frac{dz'}{dv} \right   )^2}dz'}}, \label{eq:gz}
\end{equation}
where $g_{0}$  is an arbitrary constant of integration.    It is
easy to relate the function $g(z)$ to the Hubble function $H(z)$.
In fact, using
\begin{equation}\label{eq:dt}   \frac{dz}{dt}=
-(1+z)H(z),
\end{equation}
and
\begin{equation}   {dz\over
dv}=(1+z)^{2}\frac{H(z)}{H_{0}}, \label{eq:hubble1}
\end{equation}
it follows that Eq.~(\ref{eq:gzeq}) can be
rewritten as
\begin{eqnarray}   \frac{d}{dz} \ln{g(z)} =
\frac{d}{dz}\left[\ln{(1+z)^{2}\frac{H(z)}{H_{0}} }\right],
\label{eq:hubble2}
\end{eqnarray}   so,
\begin{eqnarray}
g(z)=(1+z)^2\frac{H(z)}{H_0}, \label{eq:hubble3}
\end{eqnarray}
where we have imposed the initial condition $g(0)=1$. Let us remind
that one of the Friedman cosmological equations can be rewritten as
\begin{equation}
H(z)=H_{0}\sqrt{\Omega_m (1+z)^3 + \Omega_k
(1+z)^2 + \Omega_{\Lambda}},   \label{eq:hz}
\end{equation}
where $H_0$ is the present value of the Hubble constant,
$\Omega_m=\displaystyle{{8\pi G \varrho_m}\over 3H_0}$ is the
matter density   parameter,   $\Omega_k= -\displaystyle{kc^2\over
{R_{0}^{2}H_0}}$ is the curvature parameter and $\Omega_{\Lambda}=
\displaystyle{{\Lambda c^{2}}\over 3H_0}$ is the cosmological
constant parameter.   The omega parameters satisfy a simple
algebraic relation
\begin{equation}
\Omega_m+\Omega_k+\Omega_{\Lambda}=1. \label{eq:omega}
\end{equation}
The initial conditions (\ref{eq:rz1z2}) expressed in terms of
$H(z)$ become   \begin{eqnarray}   &&D(z_1, z) \arrowvert_{z= z_1}
= 0,\\ &&\nonumber\\ &&\frac{d}{dz} D(z_1, z) \arrowvert_{z = z_1}
= {\rm sign}(z - z_1) \frac{c}{(1+z_1) H(z_1)}.  \label{eq:Hcond}
\end{eqnarray}
\subsection{The angular diameter distance} To apply the angular
diameter distance to the realistic universe it is necessary to
take into account local inhomogeneities in  the distribution of
matter.  Unfortunately, so far an acceptable averaging procedure
for smoothing out   local inhomogeneities has not been developed
(\cite{Kraz}). Following Dyer \& Roeder ~(\cite{DR72}), we
introduce a phenomenological parameter
$\alpha=1-{\varrho_{clumps}\over <\varrho>}$ called the
clumpiness parameter, which is related to the  amount of matter in
clumps relative to the amount of matter distributed
uniformly~(\cite{DZ65,PRT99}). Therefore, Eq.~(\ref{eq:angdiam2}),
in the FRW case, can be rewritten  in the form \begin{equation}
\left ( \frac{dz}{dv} \right )^2 \frac{d^2D}{dz^2} +   \left (
\frac{d^2z}{dv^2} \right ) \frac{dD}{dz} +   \frac{3}{2} \alpha
\Omega_{m} (1+z)^5 D = 0 . \label{eq:angdiamalpha} \end{equation}
Let us note however that this equation does not fully
describe the influence of non homogeneities in the distribution of
matter on the propagation of light. It takes into account only the
fact that in locally non homogeneous universe a light beam
encounters less matter than in the FRW model with the same average
matter density. Equation (\ref{eq:angdiamalpha}) does not include
the effects of shear, which in general is non zero in a locally
non homogeneous universe (see, for example SEF Chapter 4 and 11).
The effects of shear have been estimated with the help of
numerical simulations (see, for example \cite{schwe88},
\cite{linder98}, \cite{jar91}, \cite{wasa90}) but satisfactory
analytic treatment is still lacking.

It is customary to measure
cosmological distances in units of   $c/ H_{0}$, therefore we
introduce the dimensionless angular diameter distance
 $r=DH_{0}/ c$. Using Eq.~(\ref{eq:hubble1})
and Eq.~(\ref{eq:hz}), we   finally   obtain
\begin{eqnarray}\label{eq:angdiamalpha2}   &&(1+z)\left[
\Omega_{m}(1+z)^{3}+\Omega_{k}(1+z)^{2}+\Omega_{\Lambda}\right]
{d^{2}r\over     dz^{2}}+  \\ \label{eq:stdr2}&&  \nonumber\\
&&\left({7\over
2}\Omega_{m}(1+z)^{3}+3\Omega_{k}(1+z)^{2}+2\Omega_{\Lambda}\right)
{dr\over dz}\nonumber\\
&& +{3\over 2}{\alpha}\Omega_{m}(1+z)^{2}r  = 0,
\nonumber
\end{eqnarray}
with  initial conditions
\begin{eqnarray} &&r(z)\arrowvert_{z = 0} = 0, \\ &&\nonumber \\
&&\frac{dr(z)}{dz} \arrowvert_{z = 0} = 1. \label{eq:initialcond2}
\end{eqnarray}
Eq.~(\ref{eq:angdiamalpha2}) can be cast into a different form by
using $u=1/{(1+z)}$ as an independent variable instead of $z$; we get
\begin{eqnarray}
&&u^{2}(\Omega_{m}+\Omega_{k}u+\Omega_{\Lambda}u^{3}){d^{2}r\over
du^{2}}- u({3\over 2}\Omega_{m}+\Omega_{k}u){dr\over du} \nonumber \\&& + {3\over
2}{\alpha}\Omega_{m}r=0.   \label{eq:scaledr}
\end{eqnarray}
On the other hand, using the cosmic time $t$ as an independent
variable, Eq.~(\ref{eq:angdiamalpha}) assumes the form
\begin{equation} {d^{2}r\over dt^{2}}-H(t){dr\over dt}+ 4\pi
G{\alpha} \varrho_{m}(t)r=0.
\label{eq:tempodr}
\end{equation}
This equation was for the first time introduced by Dashevski \&
Zeldovich   (\cite{DZ65}) (see also~\cite{DS66}), and more recently
Kayser et al.   (\cite{KHS97}) and ~(\cite{Kant98}) have used it to
derive an equation similar to Eq.~(\ref{eq:stdr2}). In
Eq.~(\ref{eq:tempodr}) the clumpiness parameter $\alpha$ is usually
considered as a constant. However, in (\cite{DZ65} and \cite{KHS97}),
$\alpha$ is allowed to vary with time, but only the case $\alpha=$
constant is really considered. For a discussion of the general case
when $\alpha$ depends on $z$ see, for example, the paper by
Linder~(\cite{Lin88}).\\    To give an example of our procedure of
constructing the angular diameter distance between a fictitious
observer at $z_{1}$ and a source at $z_{2}$, let us consider the
simple case of $\alpha=$ constant, ~$\Omega_{\Lambda}
=0$, and $\Omega_{k} =0$. In this   case it is easy to integrate
the equation (22), we obtain \begin{equation}
r(z)={{(1+z)^{\beta}-(1+z)^{-\beta}}\over {2\beta (1+z)^{5/4}}},
\label{eq:SEFsol}
\end{equation}
where $\beta={1\over4}\sqrt{25-24\alpha}$, is the general solution of
Eq.~(\ref{eq:stdr2}) with the imposed initial conditions. This
solution can be found, for example, in SEF (\cite{SEF}), (see
Eq.~(4.56) there). The function  $g(z)$ can be easily computed in the
flat $\Lambda=0$ Friedman-Robertson-Walker cosmological model; in
fact, in this case we have
\begin{eqnarray} H(z) = H_0 (1+z)
\sqrt{z + 1},
\end{eqnarray}
so that, according to the Eq.~(\ref{eq:hubble3}), we have
\begin{eqnarray} g(z) = (1+z)^3\sqrt{z + 1}.
\end{eqnarray}
Thus  using (\ref{eq:dzunoz}), we get the familiar solution
\begin{equation}
D(z_{l}, z_{s})=\frac{1}{2\beta}\left[\frac{(1+z_2)^{\beta-
\frac{5}{4}}}{(1+z_1)^{\beta+\frac{1}{4}}}-
\frac{(1+z_1)^{\beta-\frac{1}{4}}}{(1+z_1)^{\beta+\frac{5}{4}}}\right],
\end{equation} found in SEF.

\section{Exact solutions}
In the general form the Eq.~(\ref{eq:angdiamalpha2}) is very
complicated. From the mathematical point of view this equation is of
Fuchsian type with four regular singular points and a regular singular
point at infinity~(\cite{Ince}). General properties of this equation
have been extensively studied by Kantowski
(\cite{Kant98,Kant20,Kant01}).\\Following Kantowski let us cast the
Eq.~(\ref{eq:angdiamalpha2}) into a different form replacing $1+z=x$
and introducing $r(z)={h(z)\over {1+z}}$, we obtain
\begin{eqnarray}
\label{eq:heun1}
&&\left(\Omega_{m}x^{3}+\Omega_{k}x^{2}+\Omega_{\Lambda}\right)
{d^{2}h\over dx^{2}}+ \left({3\over 2}\Omega_{m}x+\Omega_{k}\right)x
{dh\over dx}+
\nonumber\\&&\left({3\over 2}(\alpha -1) \Omega_{m}x-\Omega_{k}\right)h=0.
\end{eqnarray}
When $\Omega_{k}\not=0$ by rescaling $x$,~
$x={\Omega_{k}\over \Omega_{m}} {\tilde x}$ this equation can be
turned into the Heun equation \begin{eqnarray}
&&({\tilde x}^{3}+{\tilde x}^{2}+ {{\Omega_{\Lambda}\Omega_{m}^{2}}\over
\Omega_{k}^{3}}) {d^{2}h\over d{\tilde x}^{2}}+({3\over 2}{\tilde x}+1){\tilde x}{dh\over
d{\tilde x}}+ \nonumber\\
&& [{3\over 2}(\alpha-1){\tilde x}-1]h=0.
\label{eq:heun2}
\end{eqnarray}
The angular diameter distance can be expressed in terms of basic solutions
of the Heun equation as
\begin{equation}
r(z)={h({\Omega_{m}\over \Omega_{k}}(1+z))\over{1+z}},
\label{eq:heun3}
\end{equation}
provided that $h({\Omega_{m}\over \Omega_{k}})=0$ and
$h'({\Omega_{m}\over \Omega_{k}})={\Omega_{k}\over \Omega_{m}}$. The
generic solution for the angular diameter distance expressed in terms
of the Heun functions is very complicated, in an explicit form it is
given by Kantowski~(\cite{Kant98}). \\When $\Omega_{k}=0$, dividing
Eq.~(\ref{eq:heun1}) by $\Omega_{m}$, we obtain \begin{eqnarray}
&&(x^{3}+{\Omega_{\Lambda}\over \Omega_{m}}){d^{2}h\over dx^{2}}+
{3\over 2}x^{2}{dh\over dx} +{3\over 2}(\alpha-1)xh=0.
\label{eq:heun4}
\end{eqnarray}
By changing variables to
$\eta(x)=\sqrt{1+\displaystyle{\Omega_{\Lambda}\over {\Omega_{m}x^{3}}}}$ and
introducing new independent variable $P(x)=x^{3/4}h(x)$ this equation
can be transformed into the associated Legendre equation
\begin{equation}
(1-\eta^{2}){d^{2}P\over d\eta^{2}}-2\eta{dP\over d\eta}- {1\over
36}(5+{{25-24\alpha}\over {(1-\eta^{2})}})P=0.
\label{eq:assL}
\end{equation}
In this case ($\Omega_{k}=0$) the angular diameter distance
is given by
\begin{eqnarray}
\label{eq:assL1} &&r(z)=
{2\Gamma(1-{\delta\over6})\Gamma(1+{\delta\over
6})\over{\delta}}(1+z)^{-{1/4}}
\sqrt{1+{\Omega_{\Lambda}\over \Omega_{m}}}{\times}\\
&&\left[P^{\delta\over 6}_{-{1\over 6}}
\left(\sqrt{1+{\Omega_{\Lambda}\over \Omega_{m}(1+z)^{3}}}\right)
P^{-{\delta \over 6}}_{-{1\over 6}}
\left(\sqrt{1+{\Omega_{\Lambda}\over \Omega_{m}}}\right)-\right.\nonumber \\
 && \left.P^{\delta\over 6}_{-{1\over 6}}
\left(\sqrt{1+{\Omega_{\Lambda}\over \Omega_{m}}}\right)
P^{-{\delta \over 6}}_{-{1\over 6}}
\left(\sqrt{1+{\Omega_{\Lambda}\over
{\Omega_{m}(1+z)^{3}}}}\right)\right]\nonumber,
 \end{eqnarray}
where $\delta=\sqrt{25 - 24\alpha}$,
and $P^{\mu}_{\nu}(z)$ denotes the associated Legendre function
of the first kind. In section 5 we will propose a simple
approximate analytic solution of the general equation for the
angular diameter distance which is reproducing reasonably well the
exact numerical solution in the range of redshifts (0, 10) far
exceeding the range of redshifts of observed supernovae of type
Ia.

\section{Angular diameter distances in the gravitational lensing theory.}
The angular diameter distances, or their combinations, appear in
the main equations and in the most important observational
quantities of the gravitational lensing (GL) theory. This fact,
together with the relation between the angular diameter distance
$(D_A)$ and the luminosity distance $(D_L)$
\begin{equation}
D_{L}=\left(1+z\right)^2D_A,
\end{equation}
makes the study of the
equation for the angular diameter distance still more important,
also from the point of view of better interpretation of
observational data. In this section we describe very briefly the
role of angular diameter distance in the gravitational lensing
theory, mentioning only some effects in which it plays an
important role.\\ Let us begin with the general expression for the
time delay between different light rays reaching the observer
\begin{equation} c \Delta t = (1+z_d) \left\{ \frac{D_d
D_s}{D_{ds}}(\vec{\theta} - \vec{\beta} )^2 -
\psi(\vec{\xi})\right \}   + {\rm constant}, \label{eq:timedelay}
\end{equation}
where $D_d$, $D_s$ and $D_{ds}$ are correspondingly the angular
diameter distances to the deflector, to the source and between the
deflector and the source. The term $(\vec{\theta} - \vec{\beta})$ in
Eq.~(\ref{eq:timedelay}) represents the geometrical time delay and the
other term is connected with the non homogeneous distribution of
matter~(\cite{SEF}). The first term in Eq.~(\ref{eq:timedelay}) is
proportional to an important combination of angular diameter distances
namely, to $\displaystyle\frac{D_d D_s}{D_{ds}}$. This combination of
angular diameter distances was introduced for the first time by
Refsdal~(\cite{Refsdal66}), and was used to obtain an estimate of
$H_0$ from the time delay measurements of multiply imaged quasars. It
is possible to rewrite it in the following way
\begin{equation}
(1+z_d) \frac{D_d D_s}{D_{ds}}=\frac{c}{H_0} [
\chi(z_d) - \chi(z_s) ]^{-1} \ ,
\label{eq:firstpiece}
\end{equation}
where the function $\chi$ is given by
\begin{eqnarray}
\label{eq:chi}
\chi(z, \Omega_m, \Omega_{\Lambda}, \Omega_k,   \alpha)
&\equiv&\int_z^{\infty}{\frac{dz}{r^2 g(z)}}=\nonumber\\&&
\int_z^{\infty}{{H_{0}\over {r^{2}(z)(1+z)^2H(z)} dz}}.
\end{eqnarray}
So, $\chi$ is connected with the general solution of
Eq.~(\ref{eq:stdr2}), and it directly appears in the expression
for time delay.\\ Let us now consider the cosmological lens
equation. The metric describing FRW cosmological models is
conformally flat. In this case the simplest way to derive the lens
equation is to use  the Fermat principle, so we have
\begin{equation}
\frac{\partial \Delta t}{\partial \vec{ \theta}}=0,
\end{equation}
or
\begin{equation}
\vec{\beta} = \vec{\theta}-\frac{2R_g}{cH_0}(1 + z_d)[ \chi(z_d) -
\chi(z_s)] \frac{\partial{\psi}}{\partial{\vec{\theta}}} \ ,
\label{eq:lenseq}
\end{equation}
where $R_g$ is the Schwarzschild radius of the deflector. By
denoting\,$\vec{\xi} = D_d \vec{\theta}$, $\vec{\eta} =
D_s\vec{\beta}$, and $\vec{\alpha}(\vec{\xi}) =
\displaystyle{\frac{2R_g}{D_d}\frac{\partial{\psi}}{\partial{\vec{\theta}}}}$,
we can transform Eq.(\ref{eq:lenseq}) into
\begin{equation}
\vec{\eta}=\frac{D_s}{D_d} \vec{\xi} - D_{ds} \vec{\alpha}(\vec{\xi}) \ ,
\label{eq:lenseq2}
\end{equation}
which is formally identical with the lens equation for $z << 1$. As is
apparent, in the lens equation (\ref{eq:lenseq2}) another
dimensionless combination   of angular diameter distances, $D_s/D_d$
appears, besides the angular diameter   distance itself, $D_{ds}$. In
other words, in the equations describing gravitational lensing we find
quantities, which can
be written in terms of the angular diameter distance $D(z)$ and the
$\chi(z)$ function. In the general case $D(z)$ and $\chi(z)$ can be
evaluated only numerically. In the next section we propose a simple
analytical approximation for both functions. We hope that these
approximate forms can be useful, for example, in big numerical codes
used to derive basic cosmological parameters from observational data
and to study for instance, statistical lensing, weak lensing,
microlensing of QSO, etc.
\section{The approximate expression for the angular diameter distance}
In the generic case the equation (\ref{eq:angdiamalpha2})
does not have analytical solutions~(\cite{Kant98,Kant20}). From the
mathematical point of view this equation is of the Fuchsian type
(\cite{Ince}) with four regular singular points and a regular singular
point at infinity. The solutions near each of the singular points,
including the point at infinity, are given by the Riemann ${\mathcal
P}$-symbol and in general the solutions can be expanded in a series of
hypergeometric functions (\cite{TRIC61}) or expressed in terms of the
Heun functions (\cite{Kant98}).\\ In practice, in the general case,
the equation (\ref{eq:angdiamalpha2}) with appropriate initial conditions
is solved
numerically. In Fig.1 we show a numerical solution of the D-R equation
for $\alpha=0.8$, $\Omega_{\Lambda}=0.7$, $\Omega_{k}=0$ and
$\Omega_{m}=0.3$.
\begin{figure}[ht]
\begin{center} \includegraphics[width=8cm, height=5.35 cm]{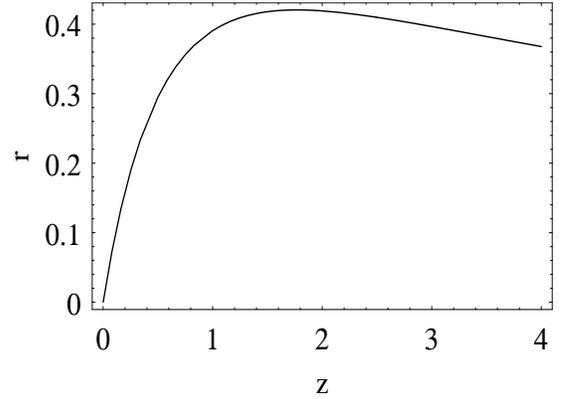} \end{center}
\caption{The dimensionless angular diameter distance as a function
of the redshift in a flat cosmological model with $\Omega_{m}=0.3$,
~$\Omega_{\Lambda}=0.7$ and $\alpha=0.8$.}
\label{fig:fig1}
\end{figure}
Using Mathematica we
discovered that there is a simple function $r(z)$ which quite accurately
reproduces the exact numerical solutions of the equation
(\ref{eq:angdiamalpha2})  for $z$ up to
10, it has the form
\begin{equation}
r(z)=\frac{z}{{\sqrt{d_1\,z^2 +
       {\left( 1 + d_2\,z +
          d_3\,z^2 \right) }^2}}},
\label{eq:approx}
\end{equation}
where $d_1$, $d_2$ and $d_3$ are constants which depend on the
parameters that specify the considered cosmological model. First of
all please note that the function (\ref{eq:approx}) automatically
satisfies the imposed initial conditions, so $r(0)=0$ and
$\displaystyle{dr \over dz}(0)=1$. To express the constants $d_1$,
$d_2$ and $d_3$ in terms of $\Omega_{\Lambda}$, $\Omega_{m}$,
$\Omega_{k}$ and $\alpha$ we inserted the proposed form of $r(z)$
into the equation (\ref{eq:angdiamalpha2}) and required that it
be satisfied to the highest order. In this way we obtained that:

\begin{eqnarray}
d_1&=&\frac{-\left(2\,\Omega_k + 3\,\Omega_m \right)^2}{4} +
\frac{7\,\Omega_k+\left( 13+ \alpha\right) \,\Omega_m}{2}-\\&&
\frac{14\,\Omega_k +\left(28 + 5\,\alpha\right)\,\Omega_m}{6\,\Omega_k +
7\,\Omega_m +4\,\Omega_{\Lambda}}, \nonumber \\ d_2&=&1 +\frac{2\,\Omega_k
+3\,\Omega_m}{4},\\ d_3&=&\frac{1}{16}\left(\left(2\,\Omega_k + 3\Omega_m
\right)^2
-4\,\left(5\,\Omega_k + 9\,\Omega_m \right)\right.+\\&&\left. \frac{8\,\left( 14\,\Omega_k
+\left(28+ 5\alpha \right) \,\Omega_m\right)}{6\,\Omega_k +7\,\Omega_m +
4\,\Omega_{\Lambda}}\right).\nonumber
\end{eqnarray}
In Fig. 2 we show the approximate solution of the equation
(\ref{eq:angdiamalpha2})  for
$\alpha=0.8$, $\Omega_{\Lambda}=0.7$, $\Omega_{k}=0$, and $\Omega_{m}=0.3$
and for comparison we also plot the exact solution.

\begin{figure}[ht]
\begin{center}
\includegraphics[width=8cm, height=5.35 cm]{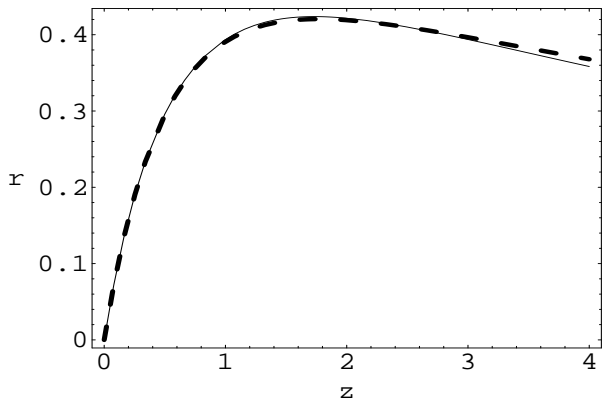}
\end{center}
\caption{The exact dimensionless angular diameter distance $r$ is
plotted (solid line) and compared with the analytic
approximation (dashed line) in a model with  $\alpha=0.8$,
$\Omega_{\Lambda}=0.7$, $\Omega_{k}=0$ and $\Omega_{m}=0.3$.
In the considered range of redshifts the relative error is
less than $10 \%$. }
\label{fig:fig}
\end{figure}
Using Mathematica we have also
found a useful approximate analytical form for the  $\chi$ function
\begin{equation}
\chi(z)={1\over{z\left(1+\displaystyle{z \over
\gamma}\right)}},
\label{chiapp}
\end{equation}
where $\gamma$ is a constant. Unfortunately we have not been able
to analytically relate $\gamma$ to $\alpha$ and other cosmological
parameters and the appropriate value of $\gamma$ should be
obtained by the standard fitting procedure. In Fig. 3 we show the
exact numerical solution for $\chi(z)$ and for comparison we plot the
approximate solution with the best fit value of $\gamma=6$.
\begin{figure}[ht]
\begin{center} \includegraphics[width=7 cm, height=7.35 cm]{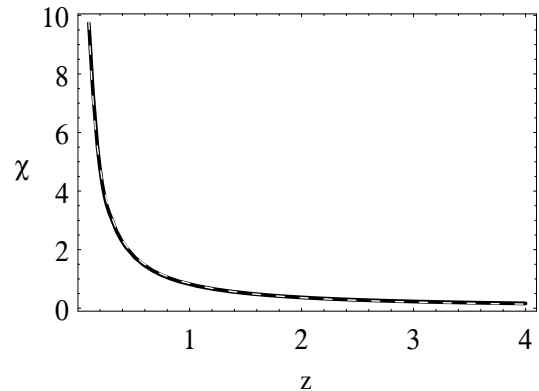}
\caption{ The exact dimensionless function  $\chi$ (solid line) for
a model with $\alpha=0.8$, $\Omega_{\Lambda}=0.7$, $\Omega_{k}=0$,
and $\Omega_{m}=0.3$. With dashed line we plot the analytic approximation
with the best fit value of the parameter $\gamma=6$.}
\label{fig:fig3}
 \end{center}
\end{figure}

\section{Conclusions}
In this paper we discuss the angular diameter distance in the
Friedman-Robertson-Walker cosmological models and consider the case
when the cosmological constant and the curvature of space could be
different from zero. The effects of local non homogeneous distribution
of matter are described by a phenomenological parameter $\alpha$,
consistently with the so called empty--beam approximation.
Unfortunately, at the moment there are no generally accepted models
that describe the distribution of baryonic and dark matter and
therefore the influence of inhomogeneities of matter distribution can
be included only at this approximate level. In the generic case
the equation
(\ref{eq:angdiamalpha2}) is of a Fuchsian type, with four regular
singular points and one regular singular point at infinity. The
general solution of this type of ordinary differential equation is
given in terms of the Heun functions~(\cite{Kant98,dem20}).

However the exact solution is so complicated that it is useless in
practical applications ~(\cite{Kant98,Kant20}). Therefore we have
proposed an approximate analytic solution, simple enough to be used in
many applications and at the same time it is
sufficiently accurate, at least in the interesting range of redshifts
($0\leq z \leq 10$). In Fig. 2 we compare the exact numerical solution
of the equation for the angular diameter distance with the approximate
one. The approximate solution
reproduces the exact curve quite well and the relative error does not
exceed $10\%$.
\\ Following SEF, we have found the function $\chi$ which appears in
the expression  for time delay as well as in the lens equation, and
which naturally appears in the expression for angular diameter
distance between two arbitrary objects at redshifts $z_1$ and $z_2$~(
see Eq.~(\ref{eq:dzunoz})).\\ We have also proposed an approximate
analytical form of the function $\chi$ which depends only on one
parameter $\gamma$ but unfortunately $\gamma$ has to be fixed by a
standard fitting procedure (see Fig. 3). Our approximations have been
already applied in complicated codes used to study the statistical
lensing (\cite{perrotta21}). \\ Finally we would like to stress that
from our analysis it follows that variations in the angular diameter
distance caused by the presence of cosmological constant are quite
similar to the effects of a non homogeneous distribution of matter
described here by the clumpiness parameter $\alpha$. In Fig. 4 we plot
the angular diameter distance for two models, one with a homogeneous
distribution of matter ($\alpha=1$) and $\Lambda\not=0$ and another
with a non homogeneous distribution   of matter~($\alpha=0.7$)
and $\Lambda=0$. We see that an inhomogeneous distribution of matter
can mimic the effect of a non zero cosmological constant. This is an
important observation in view of the recent conclusions based on
observations of high redshift type Ia supernovae that the cosmological
constant is different form zero~ (\cite{Perl97,Riess00}).

\begin{figure}[ht]
\begin{center}
\includegraphics[width=8cm, height=5.35 cm]{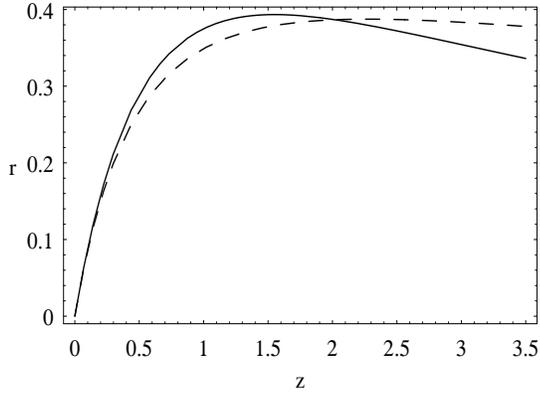}
\end{center}
\caption{The angular diameter distance
for two models, one with a homogeneous distribution of matter
($\alpha=1$) and $\Lambda\not=0$ (solid line) and another with a non
homogeneous distribution of matter ($\alpha=0.7$) and
$\Lambda=0$~ (dashed line).}
\label{fig:sn1}
\end{figure} ~\\

\section{Acknowledgments} It is a pleasure to thank
V.F. Cardone, G. Covone, C. Rubano, P. Scudellaro, and M. Sereno for
the discussions we had on the manuscript. E.P. is partially supported
by the P.R.I.N. 2000 SINTESI. This work was supported in part by EC network HPRN-CT-2000-00124. M.D. wants to acknowledge partial
support of the Polish State Committee for Scientific Research through
grant Nr. 2-P03D-017-14 and to thank the Theoretical Astrophysics
Center in Copenhagen for warm hospitality.

\section{Appendix}

As was already noted the equation for the angular diameter distance in a
locally non homogeneous Friedman-Robertson-Walker universe with non zero
cosmological constant can be reduced to the Heun equation (see Eq. 28)
(Heun, 1889). The Heun differential equation generalizes the Gauss
hypergeometric equation, it has one more finite regular singular point
(see Ince, 1964). This equation often appears in physical problems in
particular in studying of diffusion, wave propagation, heat and mass
transfer, and magnetohydrodynamics (see Ronveaux, 1995). In the general
form the Heun equation can be written as
\begin{eqnarray}
&&{d^{2}H\over dy^{2}}+
\left({a_{1}\over {y-y_{1}}}+{a_{2}\over {y-y_{2}}}+{a_{3}\over {y-y_{3}}}\right)
{dH\over dy}+\nonumber\\&&{(a_{4}y-q)\over {(y-y_{1})(y-y_{2})(y-y_{3})}}H=0,
\label{gheun}
\end{eqnarray}
where $a_{1}$, $a_{2}$, $a_{3}$, $a_{4}$, $q$, $y_{1}$, $y_{2}$ and
$y_{3}$ are constants, or in a canonical form as~(\cite{bat})
\begin{eqnarray}
&&x(x-1)(x-a){d^{2}H\over dx^{2}}+\nonumber\\
&&\left\{(\alpha+\beta+1)x^{2}-\left[\alpha+\beta+1+a(\gamma+\delta)-\delta\right]x+a\gamma\right\}
{dH \over dx}+\nonumber\\&&(\alpha\beta x -q)H=0,
\label{cheun}
\end{eqnarray}
where $\alpha$, $\beta$, $\gamma$, $\delta$, $a$, and $q$ are constants.

When $\gamma$ is not an integer, the general solution of the Heun
equation can be written as:
\begin{eqnarray}
H(x)&=&C_{1}F( a, q, \alpha, \beta, \gamma, \delta, x)+\\
&&C_{2}|x|^{1-\gamma}F( a, q_{1}, \alpha - \gamma+1, \beta-\gamma+1,
2-\gamma, \delta, x)\nonumber,
\label{exheun}
\end{eqnarray}
where $C_{1}$ and $C_{2}$ are constants and
$q_{1}=q+(\alpha-\gamma+1)(\beta-\gamma+1)-
\alpha\beta+\delta(\gamma-1)$.

Following our success in finding an approximate solution of the equation
for angular diameter distance we would like to present an approximate
solution of the Heun equation. It has the form:
\begin{eqnarray}
&&H(x)=B\frac{x}{x_0}\,\left(\frac{x}{x_0}-1 \right) \\&&
  {1\over{\sqrt{{\left( 1 + d_2\,\left(\frac{x}{x_0}-1 \right)+ d_3\,{\left(\frac{x}{x_0} -1\right)}^2 \right) }^2 +d_1\,{\left(\frac{x}{x_0} -1\right) }^2 }}\,}\nonumber,
\label{apphe}
\end{eqnarray}
where the parameter $x_0$ represents the point where the initial
conditions are specified and $d_1$, $d_2$, $d_3$, $B$ are constants.
For the Heun equation we adopt the following initial
conditions\footnote{To get the general initial conditions i.e.
$H(x_0)=A$ it is enough to add to the (\ref{apphe}) a constant
$H+A$.}:
\begin{eqnarray}\label{hecond}
 H[x_0]&=&0,\\
{dH \over dx}\left|_{x_0}\right. &=&{B\over x_0}.
\end{eqnarray}
Please note that the form (\ref{apphe}) automatically satisfies the
imposed initial conditions.

To express the constants $d_1$, $d_2$ and $d_3$ in terms of
$\alpha$, $\beta$, $\gamma$, $\delta$, $a$, and $q$, it is
necessary to insert the proposed form of $H(x)$ into the canonical
Heun equation (\ref{cheun}) and require that it be satisfied to
the highest order. The constants $d_1$, $d_2$ and $d_3$ obtained
in this way are quite complicated. It will take about 3 pages to
give them here. They are shown in an explicit form at
http://people.na.infn.it/~ester
\begin{figure}[ht]
\begin{center}
\includegraphics[width=8cm, height=5.35 cm]{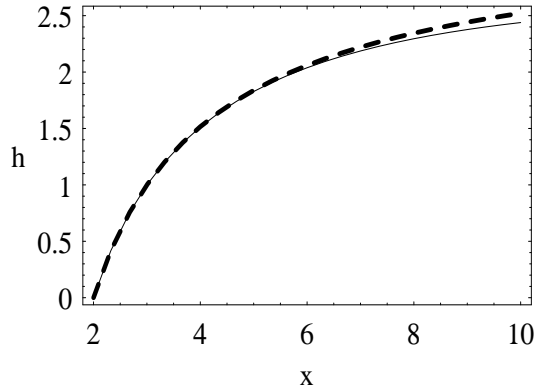}
\end{center}
\caption{Comparison of the exact (solid line) with the approximate
solution (dashed line) of the Heun equation.}
\label{fig:heun1}
\end{figure} ~
\\
\begin{figure}[ht]
\begin{center}
\includegraphics[width=8cm, height=5.35 cm]{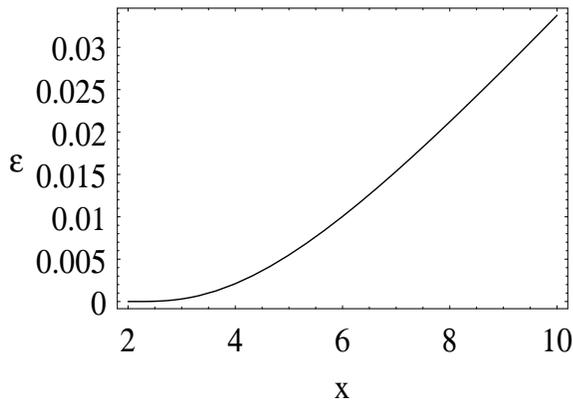}
\end{center}
\caption{Relative error of the approximate solution of the Heun
equation as a function of $x$.}
\label{fig:heun2}
\end{figure}
\\
 In Fig. 5 we compare an exact numerical solution of the Heun equation with
the approximate solution for the same values of $\alpha$, $\beta$, $\gamma$,
$\delta$, $a$ and $q$.


\newpage
\begin{thebibliography}{100}

\bibitem[Bateman \& Erd{\'{e}}lyi 1955]{bat}
Bateman, H., and Erd{\'{e}}lyi, A., {\it Higher Transcedental Functions,
Vol. 3}, 1955, McGraw-Hill Book Comp., New York

\bibitem[Carroll et al.1992]{car}
Carroll, S.M., Press, W.H.,
Turner, E.L., 1992, Ann. Rev. Astro. Astrophys.  {\bf 30}, 499


\bibitem[Dashveski \& Zeldovich 1965]{DZ65}
Dashveski V.M., Zeldovich Ya.B., 1965, Soviet Astr., {\bf8}, 854
\bibitem[Dashveski \& Slysh 1966]{DS66}
Dashveski V.M., Slysh V.I.,1966, Soviet Astr., {\bf 9}, 671
\bibitem[Dyer \& Roeder 1972]{DR72}
Dyer C.C., Roeder R.C., 1972, ApJ, {\bf 174}, L115
\bibitem[Demianski et al. 2000]{dem20} Demianski M., de Ritis
R. ,Marino A.A., Piedipalumbo E., astro-ph/0004376
\bibitem[Dyer \& Oattes 1988]{DO88}
Dyer C.C.,Oattes L.M., 1988, ApJ, {\bf 326},50
\bibitem[Dashevski \& Zeldovich 1965]{Dashzel}
 Dashevski V.M., Zeldovich Y.B., 1965,Soviet. Astr., {\bf 8}, 854.

\bibitem[Etherington 1933]{Ether33}
Etherington J.M.H., 1933, Phil.Mag., {\bf 15}, 761


\bibitem[Heun 1889]{Heun}
Heun, K., 1889, Math.Ann., {\bf 33}, 161
\bibitem[Ince 1964]{Ince} Ince E.L,
{\it Ordinary differential equation}, 1964, Dover, New York
\bibitem[Kayser et al.1997]{KHS97} Kayser R., Helbig P.,
Schramm T., 1997, A\&A, {\bf 318}, 680

\bibitem[Kayser \& Refsdal 1988]{KR88}
Kayser R., Refsdal S., 1988, A\&A, {\bf 197}, 63K
\bibitem[Kantowski 1998]{Kant98}
Kantowski R., 1998, ApJ, {\bf507},483
\bibitem[Kantowski, Kao \& Thomas 2000]{Kant20}
Kantowski R., Kao J.K., Thomas, R.C., 2000, ApJ, {\bf 545}, 549
\bibitem[Kantowski \& Thomas 2001]{Kant01}
Kantowski R., and Thomas, R.C., 2001, ApJ, {\bf 561}, 491

\bibitem[Krasinski 1997]{Kraz}
Krasinski A., {\it Inhomogeneous Cosmological Models}, 1997,
Cambridge University Press, Cambridge
\bibitem[Jaroszynski 1991]{jar91}
Jaroszynski M., 1991, MNRAS, {\bf 249}, 430
\bibitem[Linder 1988]{Lin88}
Linder E.V.,1988, A\&A, {\bf 206}, 190
\bibitem[Linder 1998]{linder98}
Linder E.V.,1998, ApJ, {\bf 497}, 28

\bibitem[Nakamura 1997]{na97}
Nakamura T., 1997, Phys. Rev. D, {\bf 40}, 2502
\bibitem[Perlmutter 1997]{Perl97}
Perlmutter S. et al., 1997, ApJ, {\bf 483}, 565
\bibitem[Perlmutter et al.1999]{perlal} Perlmutter S. et al.,
1999, ApJ, {\bf 517}, 565
\bibitem[Perrotta et al.2001]{perrotta21} Perrotta F.,
Baccigalupi C., Bartelmann M., De Zotti G., Granato G. L., 2002,
MNRAS, {\bf 329}, 445; \and astro-ph/0101430
\bibitem[Riess et al. 1998]{rei&al98} Riess et al., 1998, AJ,
{\bf 116}, 1009
\bibitem[Riess 2000]{Riess00}
Riess, A.G., 2000, PASP, {\bf112}, 1284
\bibitem[Refsdal 1966]{Refsdal66}
Refsdal S., 1966, MNRAS, {\bf132},105R
\bibitem[Refsdal 1970]{Refsdal70}
Refsdal S., 1970, ApJ, {\bf159}, 357R
\bibitem[Seitz Schneider \& Ehlers 1994]{sse94}
Seitz S.,Schneider P., Ehlers J.,  1994,  Class.Quant.Grav. {\bf 11}
2345

\bibitem[Ronveaux 1995]{ronv}
Ronveaux, A., {\it Heun's Differential Equation}, 1995, Oxford
University Press, Oxford
\bibitem[Sachs \& Kristian 1966]{Sachs}
Sachs R.K., Kristian, J., 1966, ApJ, {\bf 143}, 379
\bibitem[Schneider \& Weiss 1988]{schwe88}
Schneider P., Weiss A., 1988, ApJ,{\bf 327}, 526

\bibitem[Schneider, Ehlers \& Falco 1992]{SEF}
Schneider P., Ehlers J., Falco E.E., 1992,
{\it Gravitational lenses}, Springer\,-\,Verlag, Berlin
\bibitem[Tomita \& Futamase 1999]{PRTP99}
Tomita K., Futamase T., Gravitational lensing and the high-redshift
universe, 1999, Prog. Theor. Phys., {\bf 133}
\bibitem[Tomita et al.1999]{PRT99} Tomita K., Premadi P.,
Nakamura T., 1999  Prog. Theor. Phys., {\bf 133}, 85
%\bibitem{Tomita}{2001}{Tom1}
%Tomita K., 2000, ApJ, {\bf 529}, 38
\bibitem[Tricomi 1961]{TRIC61}
Tricomi F.G., {\it Differential equations}, 1961, Blackie, London

\bibitem[Watanabe \& Sasaki 1990]{wasa90}
Watanabe K., Sasaki M., 1990, Publ. Astron. Soc. Japan, {\bf 42}, L33
\bibitem[Weinberg 1976]{weinberg}
Weinberg S., 1976, ApJ, {\bf208}, L1
\bibitem[Weinberg 1989]{weinberg2}
Weinberg S., 1989, Rev. Mod. Phys., {\bf 61}, 1
\bibitem[Zel'dovich 1967]{zel67}
Zeldovich, Ya., B., 1967, Pis'ma Zh. Eksp. Teor. Fiz. 6, 883; (JETP
Lett. {\bf 6},316 (1967)).
\end{thebibliography}
\end{document}